\documentclass[epj]{svjour}
\usepackage{latexsym}
\usepackage{amsmath}
\usepackage{amssymb}
\usepackage{graphics}
\usepackage[dvips]{epsfig}
\begin{document}
\title{Nematic Films and Radially Anisotropic Delaunay Surfaces}
\author{Bryan Gin-ge Chen \and Randall D. Kamien 
}                     
%
%
\institute{Department of Physics and Astronomy, University of
Pennsylvania, Philadelphia PA 19104-6396}
\date{Received: date / Revised version: date}
%
\abstract{
We develop a theory of axisymmetric surfaces minimizing a combination of surface
tension and nematic elastic energies which may be suitable for
describing simple film and bubble shapes.  As a function of the elastic constant and the applied tension on the bubbles, we find the analogues of the unduloid, sphere, and nodoid in addition to other new surfaces.
\PACS{{61.30.-v}{Liquid Crystals} \and
      {82.70.Rr}{Aerosols and foams}   \and
      {47.55.D}{Drops and bubbles} 
     } 
} 
\def\makeheadbox{{}}
\maketitle
\section{Introduction and Summary}
\label{intro}
Composed of elemental air and water with a dash of soap, bubbles and
foams are fodder for studies in geometry, topology, and assembly.
Understanding their structure and dynamics is essential for many
aspects of industrial processing, flow control, and clean drains.  The description of bubbles follows geometrically simple laws: a single bubble forms a sphere to minimize its area for fixed volume.  
When multiple bubbles come into contact, their interfaces obey a set of rules,
first suggested by Plateau in 1873 \cite{plateau} and later proven by
Taylor in 1974 \cite{taylor}.  These laws, simple and universal, state
that at every edge joins three faces (at $120^\circ$) and every vertex
joins four edges (at $\cos^{-1}(-1/3)\approx 109.47^\circ$).  While the angles are set by simple force balance, the number of faces-per-edge and edges-per-vertex is fixed by area minimization -- to change these numbers (3 and 4) would require increasing the total area.   Thus, in some sense, the foam
problem is ``pure'' -- there are no coupling constants or ratios of
parameters that determine the ideal bubble shape.  In this paper we
study a simple generalization, in principle realizable  in soft
matter systems: namely, we add in-plane nematic order to the foam.
Are there more general rules for these more complicated materials or
do the shapes depend critically on a new length scale made from the
surface tension and Frank elastic constants?  We conclude that not
only are the shapes different quantitatively, but that the whole
``phase diagram'' of shapes as a function of applied tension is
dramatically changed.   The nematic order is able to support a much
richer class of shapes in addition to the classical
constant mean curvature Delaunay surfaces: the sphere, unduloid, and
nodoid.

Ultimately, we would like to predict the scaling behavior of these so-called
nematic foams with time as they coarsen, in analogy with the standard
work beginning with Lemlich for normal foams \cite{Lemlich,Markworth,PDE}.  There it
has been found that the average bubble size $\langle\,R\,\rangle \sim
t^{1/2}$, compatible with experimental results.  In preliminary experiments on liquid crystal foams
\cite{Buch} it was suggested that $\langle\,R\,\rangle \sim t^{0.2}$.
This difference from the usual coarsening exponent was suggested to be
a consequence of defects on the surfaces.  In order to study the nematic foam, we will here concentrate on 
the geometry of general
axisymmetric surfaces stationary under a sum of surface tension and nematic
elastic energies.  In a
future paper, we will consider the dynamics of a double bubble system
under diffusion \cite{TBP}. We begin by deriving a functional for this system,
then examine single bubble shapes, primarily the generalization of
axisymmetric constant mean curvature surfaces, known as Delaunay surfaces.

The functional we consider is simply that of the Frank free energy of
an axisymmetric area-minimizing surface with director field along
latitude lines:
\begin{eqnarray}
S[r(z)]=2\pi\int_0^{z_1}dz\left[\left(\gamma +\frac{K}{r^2}\right)
r\sqrt{1+\dot{r}^2} -\frac{p}{2}r^2\right]
\end{eqnarray}
We consider the minimizers of this 
functional as generalizations of Delaunay surfaces.  The Delaunay
surfaces minimize the following functional, which consists of a
surface tension term and a pressure term:
\begin{eqnarray}\label{eq:areaa}
S[r(z)]=2\pi\int_0^{z_1}dz\left[\gamma r\sqrt{1+\dot{r}^2}
-\frac{p}{2}r^2\right]
\end{eqnarray}
Minimization of (\ref{eq:areaa}) leads to the Young-Laplace law, $p=2\gamma H$, where $H$ is
the mean curvature of the interface.

Our generalization of Plateau's problem is in a different direction
than what has been studied.  Unlike the work done on surfaces in
external fields ({\sl e.g.} drops and crystals in a gravitational
field) \cite{Avron,Zia,Wente}, our anisotropy arises spontaneously
through the nematic director.  Further, our functional depends on an
internal degree of freedom and not the extrinsic geometry of the
bubble as in the Wulff model \cite{KP1,KP2}.  Previous work on fluid
membranes with in-plane order have focused on shape transitions
\cite{lubmac,lubprost} forced in by topology, not curvature.  There has been work on the Canham-Helfrich membrane model similar in spirit to our work \cite{Zh}.  Other
seminal work includes the study of nematic order on rigid membranes
\cite{VN}.  Our study of bubble shapes complements work on buckling
due to defects \cite{Frank} and vesicle shape due to tilt order
\cite{Jiang} -- there the energetics and behavior near defects was the
focus.  Here, we find a class of shapes when the curvature energy is
highly anisotropic.  

In the next section we will expand on our model and study an even more
general class of free energy functionals.  We analyze these models and
determine whether we have nodoids, unduloids, or bubbles as we vary
both the elastic constant and the applied tension along $z$.  In
section 3 we describe the actual axisymmetric shapes having reduced
their profile to tabulated functions.  In section 4 we focus on the
free bubble with no applied tension.  We conclude with section 5, the
conclusion.

\section{The Model}
\label{sec:nematicelastic}
In this section we put together the requisite energetic contributions
to derive an energy functional for a thin nematic liquid crystalline film
without any translational order.  An ordinary soap film is modeled by
a surface $S\subset \mathbb{R}^3$, which minimizes:
\begin{eqnarray}
F_{A} &=& \gamma \int_S dA
\end{eqnarray}
where $\gamma$ is the surface tension.  With a volume constraint, the
stationary points of this free energy are constant mean curvature (CMC) surfaces
which obey the Young-Laplace Law, $\Delta p = 2\gamma H$, where
$\Delta p$ is the pressure jump across the interface and $H$ is its
mean curvature.  In the absence of any volume constraints we have
$H=0$ and the surface tension does not play a role in the ground
state structure.  As Veit Elser has put it, this problem of minimal
surfaces is ``pure'' -- it does not depend on any particular
interactions or elastic constants \cite{Veit}.  Here we aim to
minimally embellish the physics of minimal and 
CMC surfaces.  We add a nematic director $\bf n$, obliged to lie in
the surface, and include its Frank free energy \cite{dG}:
\begin{eqnarray} 
F=&&\frac{1}{2}\int d^3\!x \left\{K_1\left(\nabla\cdot {\bf n}\right)^2+K_2\left[{\bf n}\cdot\left(\nabla\times
{\bf n}\right)\right]^2\right.\nonumber\\
&&\qquad\left.+K_3\left[{\bf n}\times\left(\nabla \times{\bf n}\right)\right]^2\right\} 
\end{eqnarray}
where $K_1$, $K_2$, and $K_3$ are the splay, twist, and bend elastic
constants, respectively.  When restricted to a thin film of thickness
$h\sim kT/K_i$ with planar boundary conditions on the inner and outer
surfaces, the director will be independent of the depth in the film
and we may treat it as a two-dimensional line field in the tangent plane
of the surface, $S$.

When the nematic director lies in a surface, the twist term vanishes;
without loss of generality we work in the tangent plane to an
arbitrary point so that the director is ${\bf n}=(n_x,n_y,0)$, and the
normal is in the $z$ direction.  Then
\begin{eqnarray}
{\bf n}\cdot[\nabla\times {\bf n}]&=& n_x(\partial_z n_y-\partial_y 0)
+n_y(\partial_x 0-\partial_z n_x)\nonumber\\
&&\qquad+0 (\partial_xn_y-\partial_yn_x)\nonumber\\
&=&0
\end{eqnarray}
since ${\bf n}$ is independent of $z$ at this point.  Thus the
twist term vanishes and the bend term in simply $(\nabla\times{\bf
n})^2$.

In addition to the difference between splay and bend $K_1-K_3$, the
surface normal $\bf N$ breaks the bend term into two
rotationally-invariant pieces.  Defining  ${\bf t}\equiv{\bf
n}\times\left(\nabla\times{\bf n}\right)$, we have two scalars, $t_N =
{\bf N}\cdot {\bf t}$ and $t_\perp = \vert{\bf t}\vert - t_N$.  These
two types of bend can each have a different elastic cost for bend in
the surface and out of the surface, $K_{3N}$ and $K_{3\perp}$,
respectively.  The free energy contribution from the nematic is:
\begin{equation} 
F_n=\frac{h}{2}\int_S dA\left\{K_1\left(\nabla\cdot {\bf n}\right)^2
+K_{3N}t_N^2+K_{3\perp}t_\perp^2 \right\}
\end{equation}
Note that this is, in principle, a functional of both the shape of the
surface and the director field living on it. We will focus on this
interplay in the following. 

\begin{figure}
\resizebox{8 cm}{!}
{
  \includegraphics{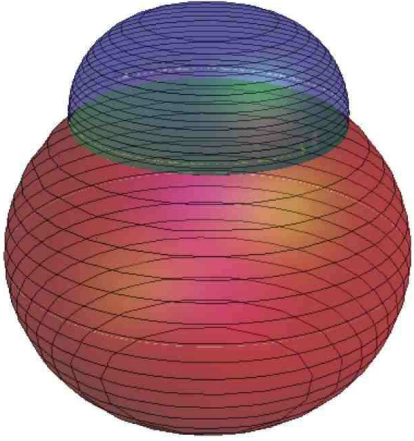}
}
\caption{Double RAD bubble, with $g(r)=1/r^2$ and $K=0.01$.  Three
$T=0$ caps meet at 120$^\circ$.}
\label{figure01}       
\end{figure}

Recall that Delaunay surfaces are CMC surfaces of revolution.
In order to generalize these for nematic films, we will focus on the
case where the director field lies parallel to the lines of
``latitude'' on a surface of revolution.  In cylindrical coordinates,
$(r,\phi,z)$, these are lines of constant $z$.  These configurations
are automatically splay-free, further removing the dependence on
$K_1$.  In future work, we will consider the fusion of two such nematic bubbles
({\sl i.e.} the double-bubble problem \cite{MorganRMP}) since this line
field is consistent with the condition that the director is parallel to the
Plateau borders (Fig. \ref{figure01}) \cite{TBP}.

In Cartesian coordinates, the director field at $(x,y,z)$ is
${\bf n}=(y,-x,0)/r$ and the normal is ${\bf N}=(x,y,z)/R$,
where $r=\sqrt{x^2+y^2}$ and $R=\sqrt{x^2+y^2+z^2}$.  Direct
calculation confirms that $\nabla\cdot {\bf n}=0$.  And 
\begin{eqnarray}
{\bf t}&=&{\bf n}\times[\nabla\times {\bf n}]\nonumber\\
&=&\frac{x \hat {\bf x} + y \hat {\bf y}}{r^2}
\end{eqnarray}
We have $|{\bf t}|^2=1/r^2$, $t_N^2=1/R^2$ and
$t_\perp^2=\left(\csc^2\theta
-1\right)/R^2$ where $\theta$ is the polar angle in polar coordinates.

In keeping with the spirit of minimally modifying the soap bubble
problem, we will set $K_{3\perp}=K_{3N}=K_3$ in the proceeding.   We
shall show that even without the anisotropies of $K_1$ and
$K_{3\perp}-K_{3N}$, the shapes of single surfaces are qualitatively
modified by $K_3\ne 0$.  In this limit, the nematic free energy only
depends on $r$, the $xy$-radius of the film so that
\begin{equation}
F_n=\frac{K_3}{2}\int dA \frac{1}{r^2}
\end{equation}
Axisymmetric surfaces may be parameterized by the $z$ coordinate so
that ${\bf R}(z)=[r(z)\cos\theta,r(z)\sin\theta,z]$.  The total free
energy as a function of the cylindrical coordinate $r(z)$
parameterizing the surface is 
\begin{equation}
F[r(z)] = 2\pi\gamma\int_0^{z_1}dz\left\{r\sqrt{1+\dot{r}^2}\left(1+\frac{K}
{r^2}\right)-\frac{p}{2}r^2\right\}
\end{equation}
where $K\equiv hK_3/\gamma$ with
has units of length-squared and the where $p=\Delta
P/\gamma$ is the reduced pressure.  Taking values for the lyotropic liquid crystal PAA,   $K_3 \sim 1.2\times10^{-6}$ dynes
and its surface tension is $\sim 35$ dynes/cm \cite{dunmur}.  Thicknesses of
typical soap films are about four microns.  Therefore,  $K\sim 1.4\times10^{-11}$ cm$^2$.   This results in a length scale
$\sqrt{K}\sim4\times10^{-6}$ cm, which is smaller than the thickness
of the film.  We point out, however, that in a dry foam, things could be quite different.  In particular, since the variation of the area is the mean curvature, $H$, we expect that after coarsening, $H$ is close to zero.  In this regime, the principal curvatures need not vanish and, indeed, our extra term is  sensitive to
them.  We will explore this possibility in a sequel \cite{TBP} which
studies the dynamics of this new class of bubbles.   In flowing polymer systems or films 
with magnetic dipoles, $K$ could be considerably larger; in the mean time, we focus on this new variational problem and show that it leads to novel shapes and geometries, even in this simple limit.

Let's take a step back from our particular model to consider a broader 
class of functionals.  We can write our energy as:
\begin{eqnarray}\label{eq:radfree}
F&=&2\pi\gamma\int dz \left\{\left[1+Kg(r)\right]r\sqrt{1+\dot{r}^2}
-\frac{p}{2}r^2\right\}
\end{eqnarray}
We interpret $1+Kg(r)$ as a spatially varying surface
energy.  This is unlike the anisotropic surface energies in the sense
of the Wulff construction for crystal shapes, in which the anisotropy
refers to a dependence on the direction of the normal vector \cite{KP1,KP2}.  This
energy is spatially varying in $r$, and has, to our knowledge, not
been considered, in contrast to the spatially varying in $z$ case \cite{Avron,Zia,Wente}.  
Note that $g(r)$ ought to be even as a function of
$r$, in order for there to be the symmetry $r\leftrightarrow -r$.  For
the case of latitudinal nematic director lines, $g(r)=1/r^2$.

We will be comparing the properties of surfaces minimizing (strictly
speaking, extremizing, as we won't study the second variation) this
free energy  with the well-studied case of Delaunay surfaces ($K=0$).
In order to give a more complete picture of the properties of surfaces
minimizing spatially varying surface tension energies, we will begin
by discussing the axisymmetric minimizers of the general family of
free energy functionals described by (\ref{eq:radfree}). We will call
these minimizers ``radially anisotropic Delaunay (RAD) surfaces''.

Since the integrand is of the form $f(r,\dot r)$, we may take
advantage of the standard first integrals of the calculus of
variations.  Because $f$ does not depend explicitly on $z$, (if $z$
were time this would lead to conservation of energy), we have:
\begin{equation}\label{eqn:forceode}
f-\dot{r}\frac{\partial f}{\partial\dot{r}}=T
\end{equation}
where $T$ is a constant.  This last equation may be interpreted as a
conservation of force along the $z$-axis, with $T$ an applied tension
or compression, depending on its sign.  For a free bubble, $T=0$,
which is the analog  of the sphere.
Setting this to other values will yield RAD surfaces analogous to the
other Delaunay surfaces, such as the nodoid and unduloid.  We shall
see that the types of RAD surfaces that are possible as we vary $T$
will depend on the value of $K$ in a nontrivial way.

If we insert our expression for $f(r,\dot{r})$, we can rewrite 
(\ref{eqn:forceode}):
\begin{equation}\label{eqn:forceode2}
\pm\gamma\frac{2\pi r}{\sqrt{1+\dot{r}^2}}\left[1+Kg(r)\right]
=T+\pi r^2\Delta P
\end{equation}
This is an equation for force balance in the $z$-direction \footnote{We note that in the study of CMC surfaces, this is an expression for
the magnitude of the weight vector \cite{KKS}, which can be used to show that CMC surfaces must asymptote to Delaunay surfaces at infinity.  Whether this notion can be generalized to our 
problem is an open question.  }: on the
right hand side, the two terms correspond respectively to the applied
tension and the pressure-induced force on a surface with $xy$
cross-section $\pi r^2$.  Likewise, the left side is the effective
surface tension $\gamma[1+Kg(r)]$ multiplied by the length of the
boundary, $2\pi r$.  The factor of $\pm(1+\dot r^2)^{-1/2}$ projects
the $z$-component of this force.  The ambiguity in sign arises from
the fact if $\dot r$ passes through infinity, the sign of the surface
tension force must change as well.  In subsequent analysis we will
need to be careful about which branch of the radical we choose, in
particular in the case of the nodoid where $\dot r$ does indeed goes
through infinity and the $z$-component of the tangent vector to the
curve changes sign.  We will cross that cut when we come to it. 

By rewriting for $\Delta P$, we can derive a generalization of the
Young-Laplace rule:
\begin{equation}\label{eq:thisone}
\Delta P=\frac{2\gamma\left[1+{K}g(r)\right]}{r\sqrt{1+\dot{r}^2}}-\frac{T}{\pi r^2}
\end{equation}
Choosing the film to be symmetric about $z=0$, we know that $\dot
r(0)=0$ and if we define $R=r(0)$, the radius of the ``waist'' then
\begin{equation}\label{eq:again}
\Delta P=\frac{2\gamma}{R}\left[1+Kg(R)\right]-\frac{T}{\pi R^2}
\end{equation}
When $K$ and $T$ both vanish, this reduces to the standard
Young-Laplace rule.  

{\sl En passant} we note that in the case of contact between two
bubbles, the effective surface tension $\gamma[1+Kg(r)]$ is {\sl
identical} for the three partial caps shown in Fig.
\ref{figure01} since the
value of $r$ is the same where they meet.  As a result, the triple
junction will join the three surfaces at $120^\circ$, just as in
regular foams.  However, the equality of surface tensions is an 
accident of axisymmetry and in general, Plateau's rules will not 
hold.

\subsection{Families of RAD surfaces}
\label{subsec:analysisRAD}

In this section we analyze the families of RAD surfaces as we change
$T$ and $K$.  First, let us measure $T$ in units of $\gamma$ as we did
for $K$ and $p$, {\sl i.e.} $\bar T=T/(\pi\gamma)$.  The force balance
equation with our radial anisotropy function specialized to
$g(r)=1/r^2$ (the nematic-derived case) now reads
\begin{eqnarray}\label{eq:fb}
\frac{r}{\pm\sqrt{1+\dot{r}^2}}\left(1+\frac{K}{r^{2}}\right)
-\frac{p}{2}r^2 =\bar T
\end{eqnarray}
If $p\neq 0$\footnote{We have not examined the $p=0$ surfaces, but
these will be to minimal surfaces as general RAD surfaces
are to CMC surfaces, and will hence be a generalization of 
catenoids.  }, we may also scale out the dependence on 
$p$ via $\tilde r=pr/2$,
$\tilde z=pz/2$, $\tilde T=p\bar T/2$, and
$\tilde K=p^2K/4$:
\begin{eqnarray}\label{eq:reduced}
\frac{\tilde r}{\pm\sqrt{1+\dot{\tilde r}^2}}\left(1
+\frac{\tilde K}{\tilde r^{2}}\right)-\tilde r^2 =\tilde T
\end{eqnarray}
In the following analyses, we will drop the tilde from our variables
unless there is any ambiguity.  There remains only one free parameter, $K$.

\begin{figure}
\resizebox{8 cm}{!}
{
  \includegraphics{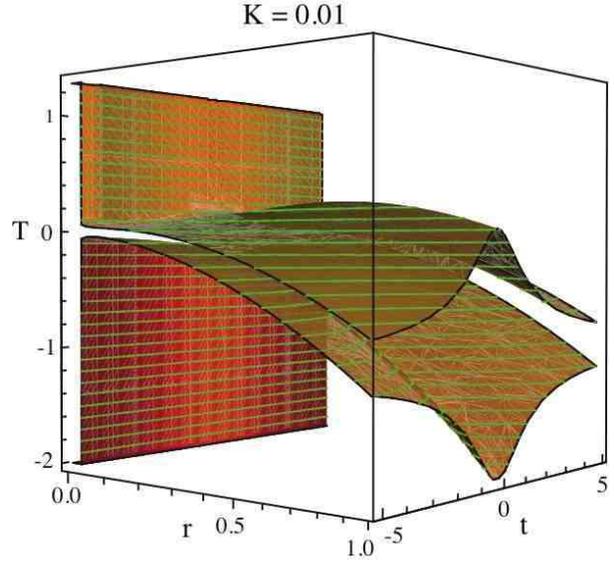}
}
\caption{The surface $T(r,\dot r)$, when $K=0.01$.  Each
connected constant height contour (green) corresponds to $r,\dot{r}$ 
for a generating curve for a RAD surface.}
\label{figure02}       
\end{figure}

We begin the study of the family of RAD surfaces with a given $K$ by
viewing $T=T(r,\dot r)$.  In Fig. \ref{figure02} we have plotted the
embedded surfaces $[r,\dot r, T(r,\dot r)]$ for both branches of the
radical.  Contours of constant $T$ correspond to values of $r,\dot{r}$
on different axisymmetric surfaces with a fixed tension.

\begin{figure}
\resizebox{8 cm}{!}
{
  \includegraphics{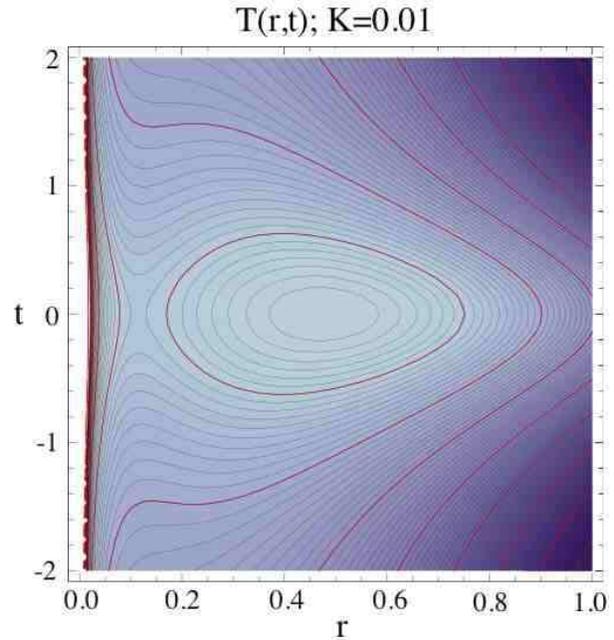}
}
\caption{The constant $T$ contours on the positive branch of $T(r,\dot r)$.  Closed contours correspond to unduloids, the single maximum is a cylindrical solution.  }
\label{figure03}       
\end{figure}

\begin{figure}
\resizebox{8 cm}{!}
{
  \includegraphics{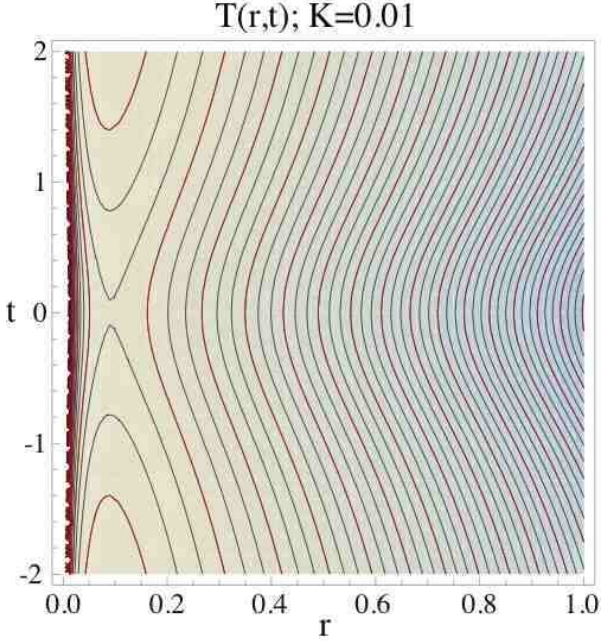}
}
\caption{The constant $T$ contours on the negative branch of $T(r,\dot
r)$.  Each contour joins with one on the positive branch as it
passes through $\dot r=\infty$.}
\label{figure04}       
\end{figure}

We start by considering the positive branch phase portrait first -- the
features sketched below should be considered as a taste of things to
come in our later analysis.  The diagram of this branch of the
$T$-surface represents points on the generating curves of RAD surfaces
where the tangent vector points upwards (has a positive
$z$-component).  In Fig. \ref{figure03} we have set $K=0.01$ and
find that there are several families of contours which cut across this
branch of the $T$-surface.  First, the $T=0$ contour represents the
curve $(r,\dot{r})$ for a free bubble, indeed, there is a range of $T$
such that the contours imply that $\dot{r}=\pm\infty$ as
$r\rightarrow0$ -- these all represent closed bubbles.  Note
that among these there are those which remain completely on the
positive branch (convex bubbles), and some which run to the negative
branch (Fig. \ref{figure04}) as well (concave bubbles).  There are
also closed curves with $T>0$ on the positive branch which represent
solutions periodic in $z$.  These are the so-called unduloidal
solutions.  The single point solution is a cylinder (since $r=$const).
Finally, there are also curves with $T<0$ which ``jump'' to the
negative branch and back, that is $\dot r=\pm\infty$ for nonzero $r$ 
-- these are the nodoids, the periodic self-intersecting surfaces,
where the tangent vector turns in a loop in each repeat unit.  Note
that not all surfaces cut the negative branch of the $T$-surface, but
all must cut the positive branch.  To see this, note that in (\ref{eq:reduced}) when $T>0$, we 
must choose the positive branch, while when $T<0$ we have solutions for both signs of the radical, in particular for large $\dot r$.  

This can be generalized for any $g(r)$.  Contours of constant $T$ are either closed (unduloids), run off to $\dot r=\pm\infty$ at $r=0$ (bubbles), or run off to $\dot r=\pm\infty$ at finite $r$ (nodoids).  Here we sketch how to partition the full space of RAD surfaces into these three classes when $g(r)$ is arbitrary by  studying the special case $g(r)=1/r^{2n}$ for $n\in\mathbb{Z}^+$.
This corresponds to assigning an energy to
the lines of latitude of $\kappa^{2n}$, where $\kappa$ is their
curvature in $\mathbb{R}^3$, and one of the principal curvatures of the surface.  We label the
second argument of $T(\cdot,\cdot)$ as $t=\dot r$ in the following and
will study the function $T(r,t)$. 

It's a fact from the theory of bifurcations in dynamical systems that the critical points of the $T(r,t)$-surface indicate where the 
geometry of the contours change, and hence those values of $T$ for which 
the geometry of the RAD surfaces change.  For example, near a local
minimum or maximum, the contours of constant $T$ are closed curves
encircling the extrema.  However, at a saddle point of the
$T$-surface, these closed contours are separated into two groups; the
separatrix is a self-intersecting curve and corresponds to a singular topological change in our surface.
Thus, if we find the critical points and
their types (minima, maxima, saddle), then each of the families
mentioned will correspond to families of contours which lie on points
of the surface between the critical points.  In order to find the
critical points, we calculate the gradient of $T$ and look for its
zeros.
\begin{eqnarray}
\nabla T&=&\pm\left(\frac{1-K(2n-1)r^{-2n}}{\sqrt{1+t^2}}-2r,
-\frac{rt(1+Kr^{-2n})}{(1+t^2)^{3/2}}\right)\nonumber
\end{eqnarray}
When $\nabla T=0$,
\begin{eqnarray}
\pm\frac{1-K(2n-1)r^{-2n}}{\sqrt{1+t^2}}&=&2r\\
{rt(1+Kr^{-2n})}&=&0
\end{eqnarray}
Since $r$ is real and positive, we must have $t=0$ and so all critical
points of $T$ must lie on the line $t=0$. The Hessian $H$
provides information on whether a critical point is a saddle, minimum,
or maximum.  At $t=0$
\begin{eqnarray}
H=\begin{pmatrix}
\pm 2Kn(2n-1)r^{-2n-1}-2 & 0\\
0 & \mp r(1+Kr^{-2n})
\end{pmatrix}
\end{eqnarray}
Since $H$ is diagonal, it is easy to see that the directions of
maximum change in $T$ are along the $r$ and $t$ directions with the
corresponding eigenvalues on the diagonal of the matrix.  Note that on
either branch, $H_{tt}$ does not change sign.  Thus on the positive
branch it is always negative and the $t=0$ line is ridge-like;
similarly on the negative branch the $t=0$ line is valley-like.  Thus
our work is simplified since we need only check $H_{rr}$ to determine
which type of extrema we have.

As we have discussed, the $T=0$ solution has significance as the
``free'' solution with no imposed tension at the boundaries, but is
also important in another respect.  Solutions with $T>0$ {\sl cannot}
have portions on the negative branch, as then both terms on the left
hand side of the force balance equation (\ref{eq:fb}) would be
negative.  This implies that the $T=0$ solution lies at the boundary
between surfaces which may be parametrized simply as $r(z)$, and those
which cannot -- a result of the behavior of the $z$-component of the
tangent vector.  Luckily, a half-repeat of any surface can be
parametrized as $z(r)$, which saves us from having to look at a
completely parametric representation $(r(s),z(s))$.

Along the ridges/valleys we have
\begin{equation}
T=\pm r\left(1+\frac{K}{r^{2n}}\right)-r^2
\end{equation}
or
\begin{equation}\label{eq:roots}
(r^{2n}+K)^2-r^{4n-2}\left(T+r^2\right)^2=0
\end{equation}
We denote $r_0$ to be the value of $r(z)$ when $dr/dz=t=0$.  All
possible values of $r_0$ are real roots of (\ref{eq:roots}).  Letting
$u=r^2$, we see that the real roots $r_j$ of the degree $4n+2$ polynomial 
in $r$ arise from positive real roots $u_j$ of the degree $2n+1$ 
polynomial in $u$
\begin{eqnarray}
P(u)&=(u^{n}+K)^2-u^{2n-1}\left(T+u\right)^2
\end{eqnarray}
On the $T$-surface, the points $(r_j,0)$ are where the contour
corresponding to a RAD surface intersects the $r$-axis.  Note that when
$T(r_j,0)$ is a critical point, $r_j$ must be a double root of
$P(r_j^2)=0$.  We will use this fact many times in what follows.
Though it would be tempting to associate each distinct positive root
$u_j$ for fixed $T$ with a distinct RAD surface, a
single generating curve may have multiple points where $dr/dz=0$.

\subsection{Reduction to Quadrature}
Indeed, while it's simple to examine the picture of the $T$-surface to
see which points are part of the same contour, the mathematics is more
subtle.  To flesh this out, we take a small detour to integrate our
force balance equation by quadratures for the function $z(r)$.  Again,
there is a sign ambiguity in the integrand, the same sign ambiguity
that we have seen throughout.  We now set the sign by the boundary
condition that $z$-component of the curve's tangent vector is positive
where we begin the integration.  We choose the origin along the
$z$-axis at the place where $\dot r=0$.  At this place $P(u)$
vanishes, by construction, and we have
\begin{eqnarray}\label{eq:z}
z(r)-0
&=&\int_{r^2(0)}^{u_1}\frac{\left(T+u\right)u^{n-1}du}
{2\sqrt{\left(u^n+K\right)^2-u^{2n-1}\left(T+u\right)^2}}
\end{eqnarray}
Note that the denominator vanishes as $\sqrt{u}$ near the lower limit and thus the singularity is
integrable.
Also, since we only integrate between consecutive roots of $P(u)$, the sign
of the integrand remains fixed.  Of course there are several roots of
$P(u)$.  Let $u_j$ for $1\leq j\leq 2k+1$ be a positive real root of
$P(u)$, with $0=u_0<u_1\leq \cdots\leq u_{2k+1}$, so that $2k+1$ is
the total number of real roots (an odd degree equation with real
coefficients is guaranteed an odd number of real roots since complex
roots must come in conjugate pairs).  The values of $T$, $K$ control
the coefficients of $P(u)$ and hence whether $k=0,1,\dots,n$.  The
integral for $z(r)$ can only remain real when $P(u)>0$ and the
intervals of $u$ such that this is true are also bounded by the $u_j$.
Therefore, only for $r^2$ such that $P(r^2)>0$ will the integral from
$u_j$ to $r^2$ result in a real valued $z(r)$.

Each interval over which $P(u)>0$ corresponds to a distinct type of
RAD surface.  When $K=0$, the only closed bubble solution is the
sphere when $T=0$.  However, when $K\neq 0$, $P(0)=K^2>0$ for any $T$
and the first valid interval is $r^2\in(0, u_1)$.  This corresponds to
closed bubbles, as $r=0$ means that the generating curve touches the
$z$-axis.  The existence of this interval for $K\neq 0$ therefore
means that there will be a closed bubble solution for any $T$.  

To exercise our formalism, we review the well-studied $K=0$ Delaunay
surfaces.  Taking $K=0$ in our equations directly yields
\begin{eqnarray}
P(u)&=(u^{n}+0)^2-u^{2n-1}\left(T+u\right)^2\nonumber\\
&=u^{2n-1}\left[u-\left(T+u\right)^2\right]
\end{eqnarray}
We will drop the factor of $u^{2n-1}$ as it yields $2n-1$ degenerate
RAD ``surfaces'' $u=0$.  Then the system is independent of $n$, and we 
recover the classical Delaunay surfaces.

\begin{figure}
\resizebox{8 cm}{!}
{
  \includegraphics{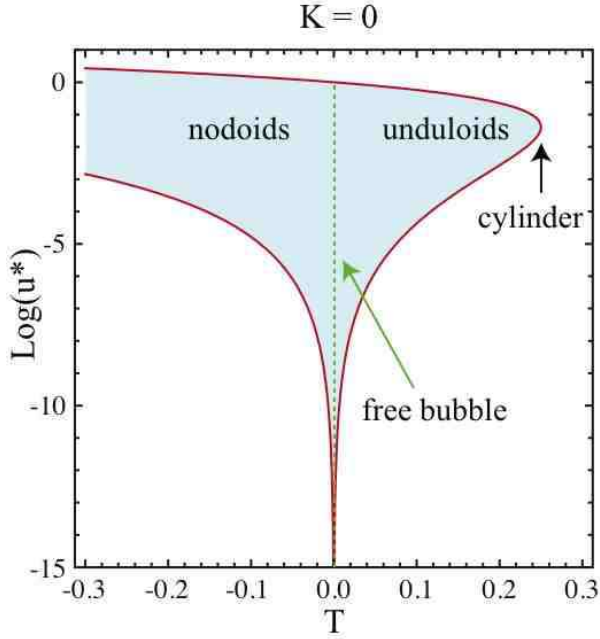}
}
\caption{A log-plot of the $t=0$ slice through the $T$-surface for 
$K=0$.  The shaded portion of the graph maps out the range of 
$u=r^2(z)$ for each Delaunay surface as a function of $T$.}
\label{figure05}       
\end{figure}

\begin{figure}
\resizebox{8 cm}{!}
{
  \includegraphics{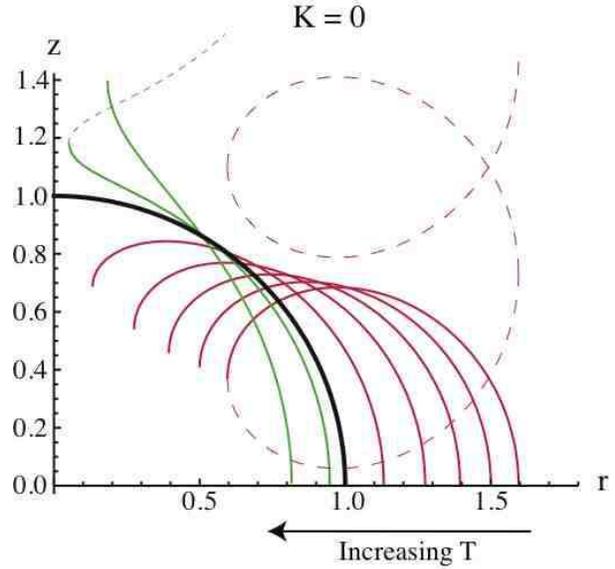}
}
\caption{Generating curves of Delaunay surfaces.  Nodoids are red,
unduloids are green, and the spherical solution 
($T=0$) is in black.  Dashed curves show a few repeats of a nodoid and an unduloid.}
\label{figure06}       
\end{figure}

\begin{figure}
\resizebox{8 cm}{!}
{
  \includegraphics{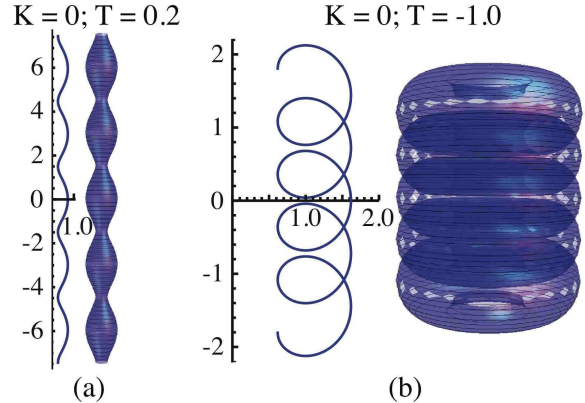}
}
\caption{Comparison of a Delaunay unduloid (a) and nodoid (b), at 
$T=0.2$ and $T=-1$ respectively.  Note that the loops formed by the self-intersections in (b) point ``inwards'', so that this is an example of an ``innie'' nodoid (c.f. Fig. \ref{figure17}d).}
\label{figure07}       
\end{figure}

In Fig. \ref{figure05} we plot $T(r,0)$ versus $\log r$; this is a
plot of the cross section of the $T$-surface through the plane $t=0$,
flipped on its side.  This graph is a ``tomographic slice'' of the
a general $T$-surface like that in Fig. \ref{figure02} except with
$K=0$.  The two branches of the
boundary curve are $T(r,0)=\pm r-r^2$ or, in other words,
$u_\pm(T)=r^2_\pm(T)$ are the solutions to the quadratic equation
$P(u)=u-(T+u)^2=0$.  The region between the two curves satisfies
$P(u)>0$, and has been shaded blue and green, for the periodic
surfaces and the closed bubble surface, respectively.  The generating
curves for Delaunay surfaces are found by setting $K=0$ in
(\ref{eq:z}), so they are described by
\begin{eqnarray}
z(r)&=&\int_{u_-}^{r^2}\frac{(T+u)du}{2\sqrt{u}\sqrt{u-(T+u)^2}}
\end{eqnarray}
where $u_-$ is the smaller root of $P(u)$
\begin{eqnarray}
u_{\pm}&=&\frac{1}{2}\left(1-2T\pm\sqrt{1-4T}\right)
\end{eqnarray}
When $T\le\frac{1}{4}$, the roots are real and $u_\pm> 0$.  Moreover,
$u_-$ vanishes only for $T=0$ and so the only closed surface at $K=0$
is the free, tensionless bubble.  One can see easily from $z(r)$ that
it is a sphere in this case.  Recall that this solution with $T=0$
separates surfaces with contours that lie entirely on the positive
branch of $T$ from those that do not.  For the Delaunay surfaces, the difference between the unduloids ($T>0$) and the
nodoids ($T<0$) is straightforward: the nodoids all self-intersect, and the unduloids do
not as shown in Figs. \ref{figure06} and \ref{figure07}.

When the two roots coincide, the integrand diverges over the
integration region unless $r^2$ is constant and equal to $u_\pm$, 
{\sl i.e.} a cylinder.   The discriminant of the quadratic equation, 
$Q(T)=1-4T$ vanishes at this point, and one can also check that $T=1/4$ 
is the only critical point on the $T(r,t)$ surface.  Since points with 
$t=0$ on the positive branch of $T$ are on a ridge, the 
cylindrical solution is the maximum of $T$. 

Finally, we may study the ``period'' of the Delaunay surfaces as we 
change $T$ as a further diagnostic of the resulting bubbles.  We define the
period as the twice the height between the endpoints of a surface or
$Z\equiv 2[z(r_+)-z(r_-)]$, where $r_i^2=u_i$:
\begin{eqnarray}
Z&=&\int_{u_-}^{u_+}\frac{(T+u)du}{\sqrt{uP(u)}}\nonumber\\
&=&\int_{u_-}^{u_+}\frac{(T+u)du}{\sqrt{u(u-u_-)(u_+-u)}}\nonumber\\
&=&\frac{2T}{r_+}K\left(\sqrt{1-\frac{u_-}{u_+}}\right)
+2r_+E\left(\sqrt{1-\frac{u_-}{u_+}}\right)
\end{eqnarray}
valid when $T\le 1/4$ and where $K(\cdot)$ and $E(\cdot)$ are the
complete elliptic integrals of the first and second kind
\footnote{\begin{eqnarray*}
F(\phi,k)&=&\int_0^\phi\frac{d\theta}{\sqrt{1-k^2\sin^2\theta}}\\
&=&\int_0^{\sin\phi}\frac{dz}{\sqrt{(1-z^2)(1-k^2z^2)}}\\
E(\phi,k)&=&\int_0^\phi\sqrt{1-k^2\sin^2\theta}d\theta\\
&=&\int_0^{\sin\phi}\frac{\sqrt{1-k^2z^2}dz}{\sqrt{1-z^2}}
\end{eqnarray*}
and $K(k)=F(\pi/2,k)$ and $E(k)=E(\pi/2,k)$}.  As shown in
Fig. \ref{figure14}, as $T\rightarrow-\infty$, the period goes to zero
(corresponding to tightly wound nodoids), and at $T=0$, $u_-/u_+=0$,
$r_+=1$ and the period, and hence the diameter of the sphere is 2.  As
$T\rightarrow1/4$, the period goes to $\pi$.  This might be
unexpected, as the period of a cylinder ought to be
undefined, or perhaps zero or infinite.  However, this result just
implies that the cylindrical limit is reached by the
vanishing of wiggles in the surface, rather than singular behavior in
the period function.

Having cut our teeth on the $K=0$ case, we move on but refer the
interested reader to other sources \cite{Kenmotsu,Eells}.  As we
mentioned, an essential difference between the $K=0$ and $K\neq0$ case
is the possibility of additional {\sl distinct} solutions for fixed
$T$.  If $K\neq0$, we can find new solutions as follows: since $P(u)>0$
in the interval $(0,u_1)$, and $P(u_1)=0$, it follows that $P(u)<0$
for $u\in(u_1,u_2)$.  Thus for $r^2\in(u_2,u_3)$, $P(r^2)>0$ and we
find a new valid solution of the force balance equation.  We may
evaluate the integral for $z(r)$ either from $r^2$ to $u_{2j+1}$ or
from $u_{2j}$ to $r^2$ -- the difference will be whether we define the
plane $z=0$ at the ``outer radius'' or at the ``inner radius'' of the
surface and corresponds only to a shift along $z$.  Continuing with
this reasoning we will find distinct surfaces by integrating values of
$r^2\in(u_{2j},u_{2j+1})$ for $j=0\ldots,k$, so that we have $k+1$
distinct surfaces.

As mentioned earlier, there is a connection between multiple roots of
$P(u)$ and critical points on the $T$-surface.
If there are repeated roots, then by perturbing $T$ slightly, we can
interpret the solutions as lying on a separatrix.  Depending on the 
parity of the colliding roots, either two intervals of positive $P(r^2)$ join 
(when $u_{2j}\rightarrow u_{2j+1}$ for some $j$) and two families of 
surfaces join into one, or an interval where $P(r^2)>0$ vanishes and the 
limits of the integral degenerate into a point
and we have an infinite cylinder of radius $r^2=u_{2j+1}=u_{2j+2}$ for
some $j$.  

Therefore, studying the discriminant of $P(u)$, which we call
$Q(K,T)$ will reveal how many positive real roots $P$ has, and hence
how many distinct RAD surfaces\footnote{Recall that for a general cubic, $ax^3+bx^2+cx+d$ with roots $r_1$, $r_2$, and 
$r_3$, we have $r_1r_2r_3=-d/a$, $r_1r_2+r_2r_3+r_1r_3=c/a$, and $r_1+r_2+r_3=-b/a$.  It follows
that $b^2c^2 - 4ac^3-4b^3d -27a^2d^2+18abcd=a^4(r_1-r_2)^2(r_1-r_3)^2(r_2-r_3)^2$.  This is the discriminant of the cubic.
}. For simplicity, we
specialize to $n=1$, where
\begin{eqnarray}\label{eq:discriminants}
P(u)&=&(u+K)^2-(T+u)^2u\nonumber\\
Q(K,T)&=&-(K-T)^2(4K+27K^2-18KT-T^2+4T^3)\nonumber\\
&=&-(K-T)^2Q_3(K,T)
\end{eqnarray}
where the last equation defines the cubic factor $Q_3(K,T)$.  Since
$P(u)$ is cubic, it may have one or three real roots. When the
discriminant vanishes, roots collide.  This may happen if $K=T$ or at a
root of $Q_3$.  $K=T$ implies that the roots of $P(u)$ are 
$\{1,-K,-K\}$.  When $K>0$ the repeated root corresponds to 
non-physical, imaginary values
of $r$ and so the vanishing of $Q(T,T)$ is of no particular physical
consequence there.  We will discuss the $K<0$ case only briefly at the 
end of this section.  We thus focus on the roots of $Q_3(K,T)$ which are
precisely the critical points in $T$, the values of tension at which
there is a crossover between a system where there is one bubble (one
real root), and one where we have a bubble and another surface (three
real roots).  We'll discuss the geometry of these surfaces later.

Moreover, the number and nature of
the critical points of $T$ varies as we changed $K$.  Note that as $K$
changes,  the positions of the roots of $Q_3(K,T)$ move as well.
Hence we may expect that if we tune $K$ through certain critical values,
the critical values of $T$ will themselves collide or split, and so we 
gain or lose entire families of RAD surfaces.  It will therefore be of use 
to consider the discriminant of $Q_3(K,T)$.  $K=0$ will necessarily be 
a critical value in $K$ because there the stationary surfaces are
simply the ordinary Delaunay surfaces along with multiple degenerate 
$u=0$ roots in $P(u)$ which we dropped in our discussion above.  Since 
this case is understood, we focus on nonvanishing 
critical values of $K$.

Since the relevant factor of the discriminant $Q(K,T)$ is cubic, we
may have one or three (real) critical points in $T$.  The discriminant
of $Q_3(K,T)$ is
\begin{eqnarray}
D_3=-16K(-1+27K)^3
\end{eqnarray}
Descartes rule of signs tells us that $Q_3(K,T)$ will always have one
(real) negative root in $T$, which we label $T_1$.  If the other two
roots are complex, then let $T_2=T_3^*=\mu+i\nu$.  Since
$D_3=64(T_1-T_2)^2(T_1-T_3)^2(T_2-T_3)^2$ we see that
in the case of one real root
$D_3=-256\left[(T_1-\mu)^2+\nu^2\right]^2\nu^2<0$.  When all three roots
are real, however, $D_3$ is real and positive.  Thus when $0<K<1/27$
there are three critical points, but for larger $K$ there is only one
critical point on the $T$-surface.

\begin{figure}
\resizebox{8 cm}{!}
{
  \includegraphics{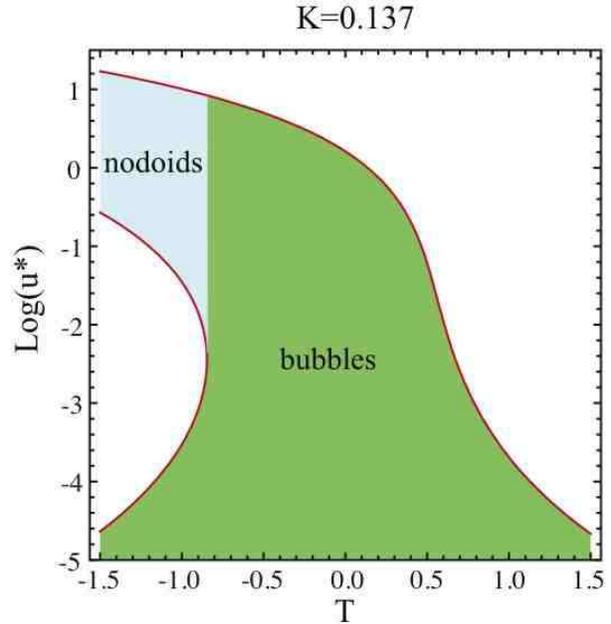}
}
\caption{The $t=0$ slice through the $T$-surface at $K=0.137$. Only
two regimes are observed here; nodoids and bubbles for $T<T_1$ and
bubbles only for $T>T_1$.}
\label{figure08}       
\end{figure}

\begin{figure}
\resizebox{8 cm}{!}
{
  \includegraphics{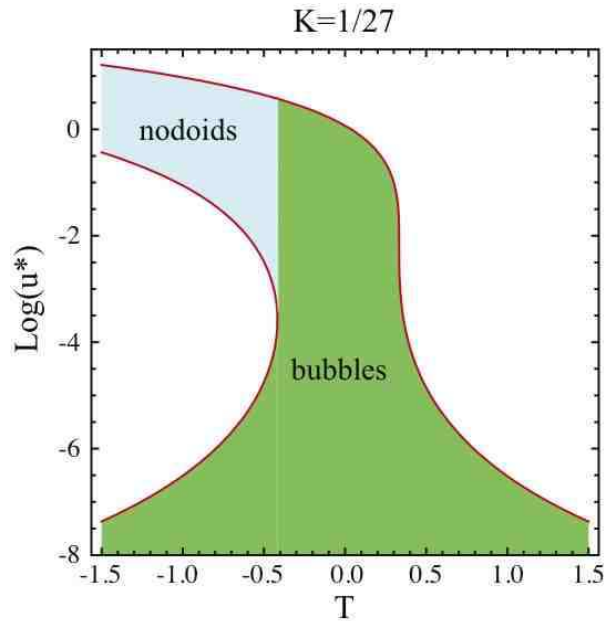}
}
\caption{The $t=0$ slice through the $T$-surface at the critical point
of $K=1/27$.  Compare Figs. \ref{figure08} ($K>1/27$) and
\ref{figure10} ($K<1/27$).}
\label{figure09}       
\end{figure}

Having deduced so much with so little, let's unravel what this implies
for $P(u)$ in each case.  We begin with the case $K>1/27$, as $T_1<0$
is our only critical point and the situation is somewhat simpler.
When $T<T_1$, there are three real roots and we will have two
intervals of positivity, $(0,u_1)$ and $(u_2,u_3)$, and hence we have
two distinct RAD surfaces, a closed bubble, which we call a ``tiny
bubble'' and another surface we call a nodoid in analogy with the
Delaunay surface since $T<0$.  $T_1$ corresponds to the point where
the nodoids join with the tiny bubble ({\sl i.e.} $u_1$ collides with
$u_2$ and they move off the real axis). When $T>T_1$ there is one real
root of $u$ and we have one interval for which $P(u)>0$, and the only
valid integration interval is the one corresponding to the closed
bubble, namely $(0,u_1)$. We note that this root is a continuation of
$u_3$ from when $T<T_1$.   In Fig. \ref{figure08}, $K=1/27+0.1$, we show a
semilog plot of the $t=0$ slice of the $T$-surface depicting the
space of surfaces. 

\begin{figure}
\resizebox{8 cm}{!}
{
  \includegraphics{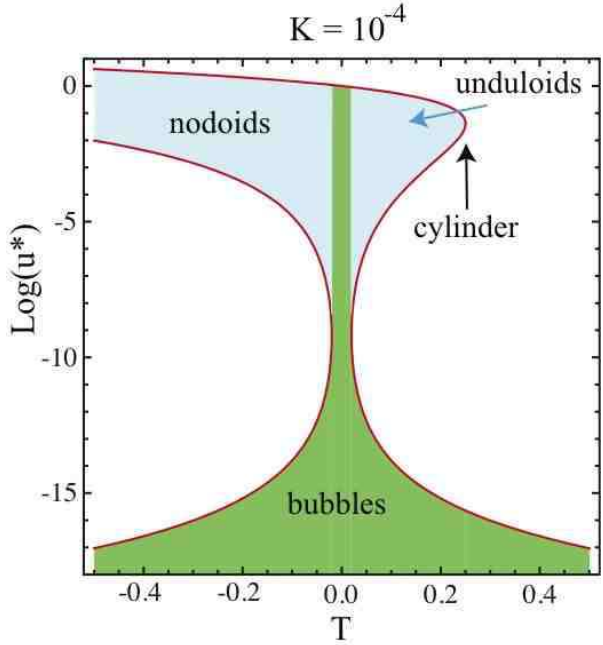}
}
\caption{The $t=0$ slice through the $T$-surface  
with $K=10^{-4}$ (qualitatively similar to Fig. \ref{figure02}, where
$K=10^{-2}$).  There are four regimes, nodoids and bubbles for $T<T_1$, 
bubbles only for $T_1<T<T_2$, unduloids and bubbles for $T_2<T<T_3$, and 
bubbles only for $T>T_3$.}
\label{figure10}       
\end{figure}

\begin{figure}
\resizebox{8 cm}{!}
{
  \includegraphics{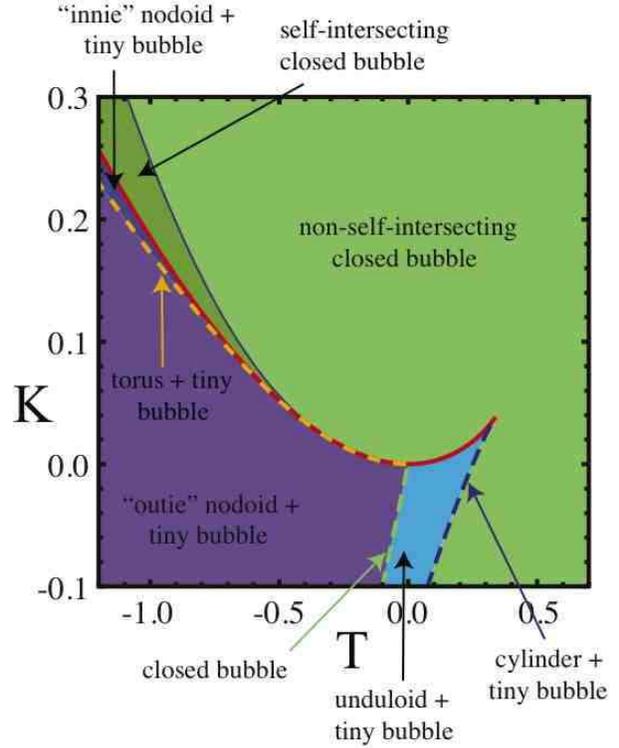}
}
\caption{RAD surface phase diagram, when $g(r)=1/r^2$.  The solid red
curve and dotted dark blue curves separate regions with one RAD surface
from regions with two RAD surfaces and are the curves defined by
the vanishing of $Q_3(K,T)$ (eqn. \ref{eq:discriminants}).  The period / height function for the surfaces diverges to $+\infty$ for $T<0$ and $-\infty$ for $T>0$ on the solid red line.  The dotted 
green line is the line $T=K$.  Together these curves are where the 
discriminant $Q(K,T)=0$.  The solid blue and dotted orange lines are 
calculated from the vanishing of the height $Z_1(T,K)$ and 
period $Z_3(T,K)$ (eqns. \ref{eq:z1},\ref{eq:z3}), respectively.  Note that the non-self-intersecting closed bubble solutions can be divided into bubbles convex at $r=0$ ($T\geq 0$) and concave at $r=0$ ($T<0$).}
\label{figure11}       
\end{figure}

On the other hand, if there are three critical points on the
$T$-surface (when $0<K<1/27$), things become more complicated.  Let
the three critical values be then $T_1<0<T_2<T_3$.  When $T<T_1$,
$P(u)$ has three real roots, and so we have two distinct RAD surfaces,
a ``tiny bubble'' and a nodoid again.  $T_1$ corresponds to the point
where the nodoids join with the tiny bubble.  For $T_1<T<T_2$, we have
only a closed bubble solution, as now $P(u)$ has only one real root.
We'll sometimes call this the large bubble, to distinguish it from the
tiny bubble solutions, though we will see (Fig. \ref{figure11}) that
they may be taken to each other via a non-singular path surrounding
the point $(T=1/3,K=1/27)$.

We have seen that for $T<0$, there is always a portion of the contour on 
the negative branch of the $T$-surface, so the large bubble with $T<0$
must be concave at $r=0$, which we can verify by calculating 
$d^2z/dr^2$ at $r=0$.  For $T\geq0$, the bubble is convex, and $T=0$
is the free bubble case which we will consider in much more detail
in section 4.  In that case, we must consider $d^4z/dr^4$ at $r=0$ to establish
convexity.

For $T_2<T<T_3$, $P(u)$ has three real roots again, and we have both
tiny bubbles and periodic non-self-intersecting RAD surfaces analogous
to unduloids.  At $T_3$, the unduloids degenerate to a cylinder,
though the tiny bubble is still there, and past $T_3$ there is only a
tiny bubble solution.  In Fig. \ref{figure10} 
we show a semilog plot of the $t=0$ slice of the $T$-surface for this
case.  Fig. \ref{figure09} shows the limiting case $K=1/27$, which
behaves most like the $K>1/27$ case.  Since $T_2=T_3$, the unduloids 
in the $0<K<1/27$ case are not visible.  Therefore the change in
character of the RAD surfaces as a function of $T$ is controlled only 
by the critical point at $T_1$.

In Fig. \ref{figure11} we show a ``phase diagram'' in $T$ and
$K$ of the types of surfaces possible.  From this, we can see the
``critical points'' at $(T,K)=(1/3,1/27)$ and $(0,0)$ are where
certain ``first order'' (branch cut) lines end.  We conclude this section with
a few sentences on the unphysical but mathematically enlightening case of 
$K<0$.  When $K<0$, $Q_3$ only has one real root $T_1$ (actually the
continuation of the root $T_3$ when $K>0$), which lies
between the regime of unduloids and tiny bubbles and the regime of
closed bubbles.  However, now the $(K-T)$ factor of the complete 
discriminant $Q$ of $P(u)$ becomes relevant.  It defines a curious 
boundary between the nodoid and the unduloid.  Though the line $T=K$
separates two regions for which there are two RAD surface solutions, the 
tiny bubble and a periodic surface (nodoid when $T<K$, unduloid on the 
$T>K$ side), on the line itself, the tiny bubble caps off the periodic 
surface to form only one solution, a larger closed bubble.  This 
behavior generalizes the nodoid to sphere to unduloid transition at 
$T=0$ for the Delaunay surfaces.  This observation thus unifies
the periodic surfaces somewhat.  Below we will again concentrate on 
surfaces with $K>0$, though our formulas are applicable to all $K$.

\section{Shapes}
We finally turn to the particulars of the allowed shapes as we explore the contours of constant $T$.
When $n=1$ the radicand in (\ref{eq:z})  is cubic in $u$ and thus the integral may be evaluated in terms of elliptic functions:
\begin{eqnarray}
z_3(r)&=&\int_{r^2}^{u_3}\frac{(T+u)du}{2\sqrt{P(u)}}\\
&=&\frac{T+u_1}{\sqrt{u_3-u_1}}F\left(\sqrt{\frac{u_3-r^2}{u_3-u_2}},
\sqrt{\frac{u_3-u_2}{u_3-u_1}}\right)\nonumber\\
&&\qquad+\sqrt{u_3-u_1}E\left(\sqrt{\frac{u_3-r^2}{u_3-u_2}},
\sqrt{\frac{u_3-u_2}{u_3-u_1}}\right)\nonumber
\end{eqnarray}
where $F(\phi,k)$ and $E(\phi,k)$ are the elliptic integrals of the
first and second kind, respectively, and the $u_i$ are the roots of
$P$.  The domain of this function is $u_2<r^2<u_3$ when we are looking
at nodoids and unduloids.  For the large bubble case, $P(\tilde{u})$ has only
one real root $\tilde{u}_1$, so we must break this naming convention when 
interpreting the formula above by taking $u_1,u_2$ to be the complex
conjugate roots and $u_3$ to be the real root. In this case, the
domain is $0<r^2<\tilde{u}_1$.  The 
tiny bubble shape can be expressed in this form as well:
\begin{eqnarray}
z_1(r)&=&\int_{r^2}^{u_1}\frac{(T+u)du}{2\sqrt{P(u)}}\\
&=&\frac{T+u_1}{\sqrt{u_3-u_1}}F\left(\sqrt{\frac{u_3-r^2}{u_2-r^2}},
\sqrt{\frac{u_3^{}-u_2}{u_3^{}-u_1}}\right)\nonumber\\
&&\qquad+\sqrt{u_3-u_1}E\left(\sqrt{\frac{u_3-r^2}{u_2-r^2}},
\sqrt{\frac{u_3-u_2}{u_3-u_1}}\right)\nonumber\\
&&\qquad-\sqrt{\frac{(u_3-r^2)(u_1-r^2)}{u_2-r^2}}\nonumber
\end{eqnarray}
Here $0<r^2<u_1$.  The integral could describe the
closed bubble as well, but our choices of elliptic integrals here turn 
out to be on the wrong branch for that case.
For large $|T|$, the root of $P(u)$ is $u_1\sim\frac{K^2}{T^2-2K}\sim
K^2/T^2$.  Since in this limit $u_1$ is small, the remainder of the
radicand varies slowly in $u\in[0,u_1]$.   Hence in this limit the bubble shape becomes spherical, 
$z_1(r)\approx -\frac{T}{|T|}
\sqrt{u_1-r^2}$; the tiny and closed bubbles become spherical
as $T\rightarrow\infty$.

\begin{figure}
\resizebox{8 cm}{!}
{
  \includegraphics{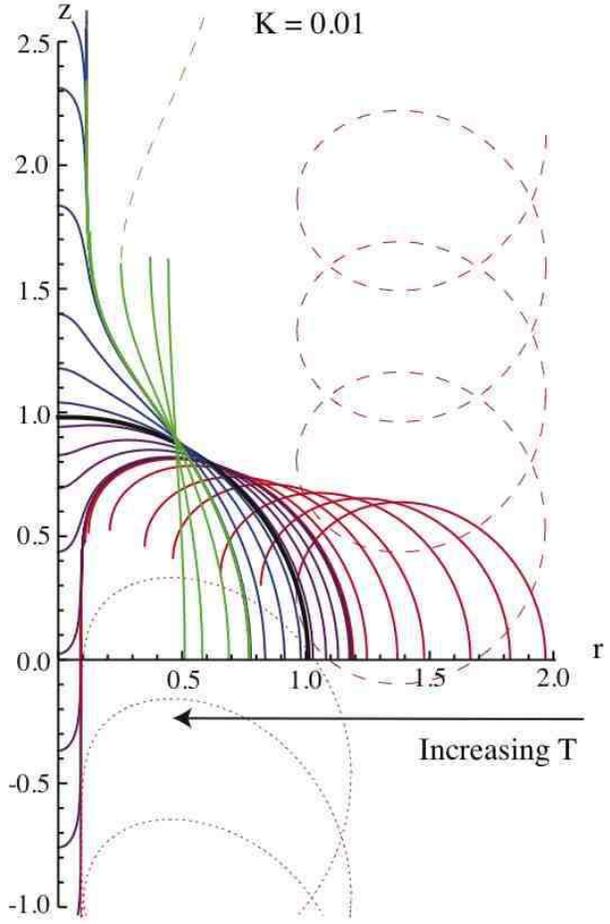}
}
\caption{Generating curves (for the nodoids, large
bubbles, and unduloids) at $K=0.01$.  The surfaces are color-coded:
nodoids are red, concave closed bubbles are purple, $T=0$ is black,
convex (at $r=0$) bubbles are blue, and unduloids are green.  Dashed curves show a few repeats of an ``innie'', an ``outie'', and an unduloid.}
\label{figure12}       
\end{figure}

\begin{figure}
\resizebox{8 cm}{!}
{
  \includegraphics{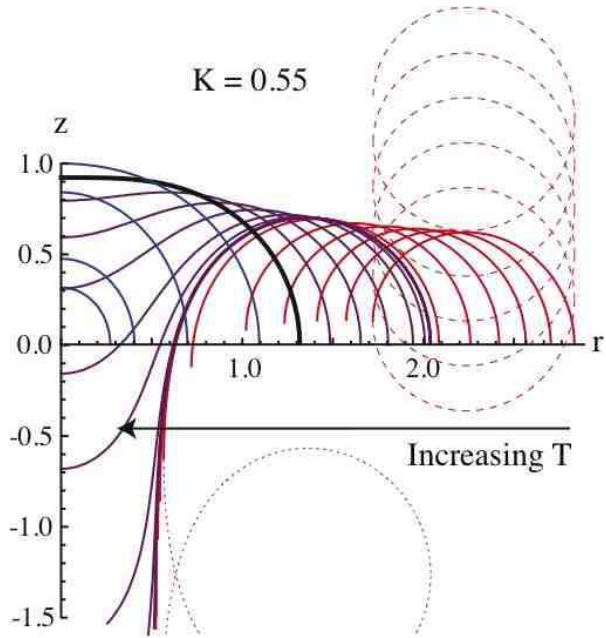}
}
\caption{Generating curves (for the nodoids and 
bubbles) at $K=0.55$.  Nodoids are red, concave closed bubbles 
are purple, $T=0$ is black, convex (at $r=0$) bubbles are blue.  Dashed curves show a few repeats of an ``innie'' and an ``outie''.}
\label{figure13}       
\end{figure}

In Fig. \ref{figure12} we show some
representative generating curves in the case $0<K<1/27$, specifically
with $K=0.01$.   Note that as $T$ increases we go from branches of nodoids which begin at $z=0$ and end at some $z\neq 0$, to large bubbles which end at $r=0$, and finally to branches of unduloids which also end at $z\neq 0$. The generating curves for $K=0.55>1/27$ are depicted
in Fig. \ref{figure13} where we transition from nodoids to concave to convex bubbles as $T$ grows.
We may also analyze the behavior of the solutions as a function of $T$ 
by considering the period function as we did for Delaunay
surfaces:
\begin{eqnarray}\label{eq:z3}
Z_3&=&\int_{u_2}^{u_3}\frac{(T+u)du}{\sqrt{(u-u_1)(u-u_2)(u_3-u)}}\\\nonumber
&=&2\frac{T+u_1}{\sqrt{u_3-u_1}}K\left(\sqrt{\frac{u_3-u_1}{u_3-u_2}}
\right)\\\nonumber
&&\qquad+2\sqrt{u_3-u_1}E\left(\sqrt{\frac{u_3-u_1}{u_3-u_2}}\right)\nonumber
\end{eqnarray}
$Z_3$ only yields the periods for nodoids and unduloids, 
and the complete elliptic integrals are chosen so that their moduli 
satisfy $0<k<1$ when $P(u)$ has three real roots, so that the integral 
is real.  For the tiny bubble, we have:
\begin{eqnarray}\label{eq:z1}
Z_1&=&\int_{0}^{u_1}\frac{(T+u)du}{\sqrt{P(u)}}=2z_1(0)
\end{eqnarray}
In the case of three roots $Z_1$ also gives the height of the closed
bubble.  In the large $|T|$ limit, since these bubbles become
spherical, $Z_1\sim2r_1=2\sqrt{u_1}\sim-2T/K$, with the sign chosen to
agree with that of $Z_1$.  The notations for $Z_j$ and $z_k$ are 
independent of each 
other, with $j$ referring to the number of real roots required to
exist for $Z_j$ to make sense, and $k$ referring to the upper limit 
of integration $u_k$.  

\begin{figure*}
\resizebox{18 cm}{!}
{
  \includegraphics{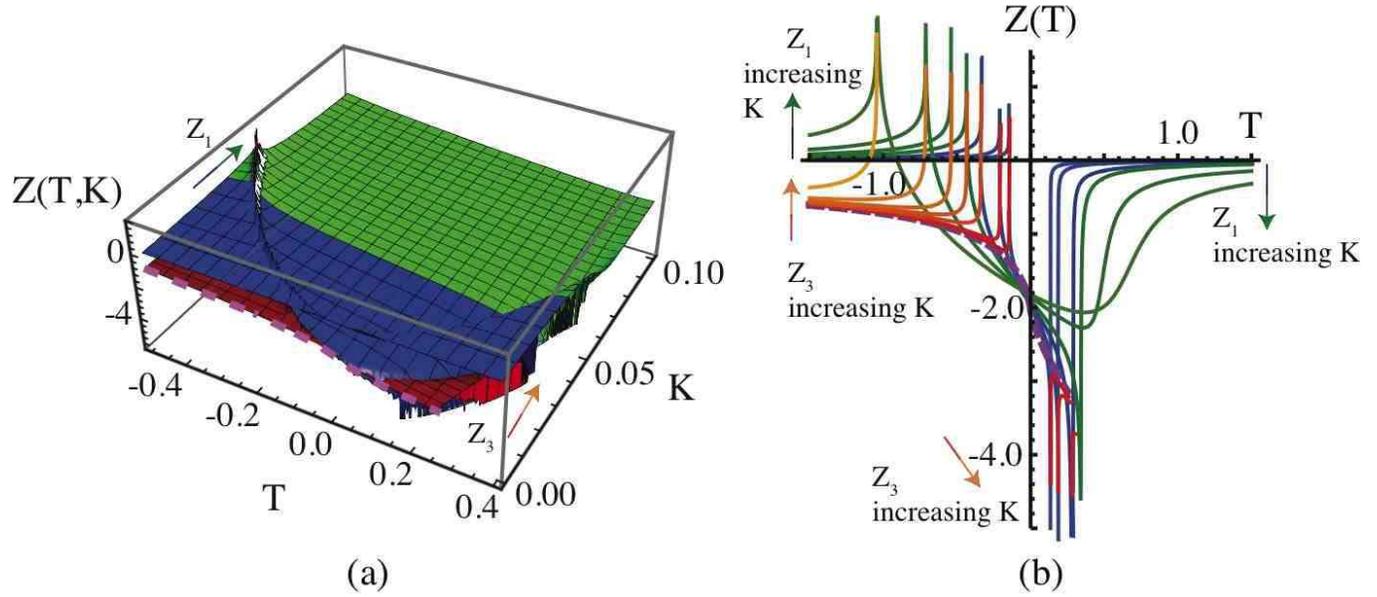}
}
\caption{The functions $Z_3$ and $Z_1$, which give the period of
unduloids and nodoids and the height of closed bubbles,
respectively. (a) The period surface $Z(T,K)$.  The $Z_3$ surface is red and the $Z_1$ 
surface with $K<1/27$ is blue, and $Z_3$ with $K>1/27$ is green.   The single period
function $Z$ at $K=0$ is the dashed and purple edge of the red $Z_3$ surface.  The 
phase diagram (Fig. \ref{figure11}) gives a ``top view'' of this
surface.  (b) Constant $K$ slices through the period surface.  Again, $Z_3$ curves are red and orange, $Z_1$ curves are blue for $K<1/27$ and green for $K>1/27$, and $K=0$ is the dashed purple curve.}
\label{figure14}       
\end{figure*}

\begin{figure}
\resizebox{8 cm}{!}
{
  \includegraphics{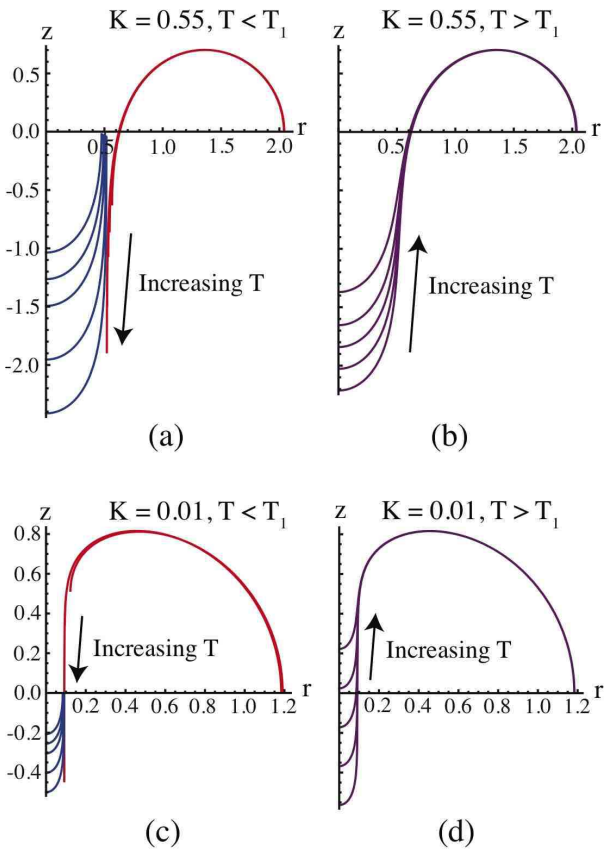}
}
\caption{The transition between nodoid and tiny bubble to large bubble
for $K=0.01$ ((a) and (b)) and $K=0.55$ ((c) and (d)).  The period of
the nodoid and height of the bubbles diverge when the two surfaces 
join.}
\label{figure15}       
\end{figure}

In Fig. \ref{figure14} we show the behavior of $Z_3$ and $Z_1$ 
for various values of $K$.  The $K>0$ period functions $Z_3$ and
$Z_1$ both have divergences at $T=T_1$ and $T=T_2$; this is visible in
the plot as vertical lines where three curves $Z_3(T<T_j)$, 
$Z_1(T<T_j)$, and $Z_1(T>T_j)$ join.  On the phase diagram (Fig.
\ref{figure11}), the divergences lie on the solid red line; the
divergence to $+\infty$ at $T_1<0$ and $-\infty$ at $T_2>0$.
Physically, this signals a loss of mechanical stability in these 
systems, as an ``elastic modulus'' of $dF_z/dz=dT/dZ$ goes to zero.
From our discussion earlier,
$T=T_1$ is where the nodoid and tiny bubble solutions merge.  Fig. 
\ref{figure15} illustrates this for $K=0.01$ and 
$K=0.55$.  We can also think of this process as the tiny bubble 
``tearing off'' as we increase the compression on a large bubble 
(decrease $T$) past $T_1$.

The nodoids close to $T_1$ have the ``loops outside'' as shown in Fig.
\ref{figure17}d, as opposed to the usual ``loop inside'' 
nodoids in the Delaunay surface case shown in Fig. \ref{figure07}b.  In 
terms of $Z_3$, we see that ``innies'' are when $Z_3<0$ and ``outies''
are when $Z_3>0$.  The period diverges at $T=T_1$ in $Z_3$, which
appears for all $K>0$ but not $K=0$.  The fact that the divergence is
positive guarantees a value of $T_{torus}<T_1<0$ where $Z_3=0$ and a torus
appears.   We show the tori at $K=0.01$ and $K=0.55$ in Figs.
\ref{figure17}e,f.  The curve $T_{torus}(K)$ is also drawn on the phase
diagram (Fig. \ref{figure11}) as well.  Note that $T_{torus}$ for $K=0.01$ is very close to
$T_1$ and the generating curve for the $K=0.55$
torus is quite close to circular.  In fact, we find numerically that 
as $K\rightarrow\infty$, $T_{torus}\sim-3K-2/3$ and the
generating curve approaches 
$\pm\sqrt{\frac{4}{9}-\left(r-\sqrt{3K}\right)^2}$,  a
circular torus with ring radius $\sqrt{3K}$ and tube radius $2/3$.  
These length scales can be derived analytically by finding the first few terms in a 
series solution (in $\sqrt{K}$) for $u_2$ and $u_3$, assuming the empirically derived behavior of$T_{torus}$. 

\begin{figure}
\resizebox{8 cm}{!}
{
  \includegraphics{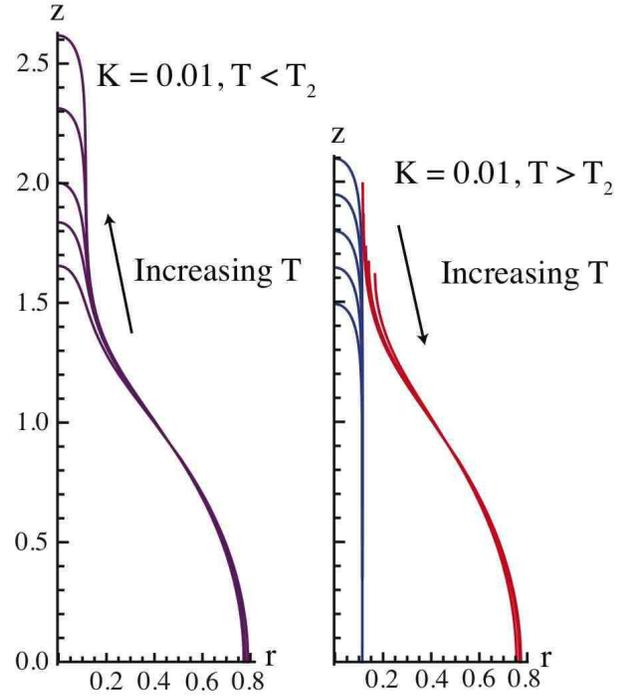}
}
\caption{The transition between large bubble to unduloid and tiny bubble
for $K=0.01$.  Note the divergence again in the heights of the
surfaces.}
\label{figure16}       
\end{figure}

\begin{figure*}
\resizebox{18 cm}{!}
{
  \includegraphics{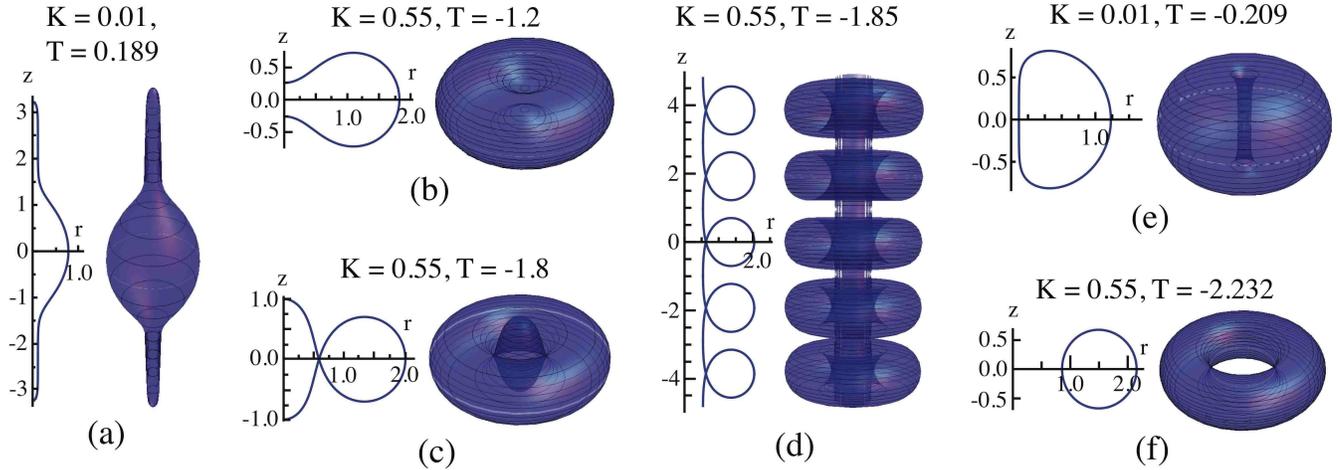}
}
\caption{From left to right: (a) A closed, protruding bubble at
$K=0.01$ with $T_2-10^{-7}$, where $T_2=0.1887$.  Concave
bubbles at $K=0.55$ with $T=-1.2$ (b) and $T=-1.8$ (c) . (d) An
``outie'' nodoid at $K=0.55$ with $T=-1.85$; $T_1=-1.848$ at this $K$.
Tori at $K=0.01$ (e) and $K=0.55$ (f).  For $K=0.01$,
$T_{torus}=T_1-4.48\times10^{-8}$ where $T_1\approx-0.209$ and for
$K=0.55$, $T_{torus}=-2.232$ and $T_1=-1.848$.}
\label{figure17}       
\end{figure*}

Finally, we turn to the closed bubble solutions: as $T$ grows
from negative to positive, the bubbles change from concave to convex
at $r=0$, this follows from the expansion $z_1(r)=\frac{T}{2K}(u_1-r^2)+\mathcal{O}(r^4)$.  
From the divergence in $Z_1$ at $T_1$ and $T_2$, we know
that the height of the bubble goes to infinity when it joins with the
nodoid and unduloid.  Closed bubbles with $Z_1=0$ are tangent to
themselves at the origin, and so $Z_1(T,K)=0$ forms the line between
self-intersecting and non-self-intersecting bubbles (Fig.
\ref{figure11}).  We show concave bubbles
with $K=0.55$ in Figs. \ref{figure17}b,c, and a bubble
close to $T_2$ with $K=0.01$ in Fig. \ref{figure17}a.   The splitting of
such a protruding bubble into a tiny bubble and an unduloid is
depicted in Fig. \ref{figure16} when $K=0.01$.  We add that this shape is reminiscent of those found in Canham-Helfrich membrane models for vesicles \cite{powerstether}. 

Note that although the unduloids end at $T=T_3$, there is no indication in 
the period function; indeed, as with the pure Delaunay
case, the cylindrical solution actually has a finite period as a limit of 
unduloids.  Neither of the elliptic integrals making up the period 
function has any divergence at this point.  Hence, in contrast to the
behavior at $T_1$ and $T_2$, the tiny bubble solution at $T_3$ is
completely regular.

\section{Free Bubbles}
\label{subsec:shapesolve}

From this point forward we discuss only the $T=0$ case of free
bubbles.  In future work we will start from these equilibrium shapes to
study the dynamics of the nematic double bubble \cite{TBP}.    As
before, we define $z=0$ at the ``equator'', where $dr/dz=0$.  We
define $R=r(0)$ as the waist of the bubble.  Specialized to $T=0$ and
$g(r)=1/r^2$, (\ref{eq:again}) reads
\begin{eqnarray}
p&=&\frac{2}{R}\left(1+\frac{K}{R^2}\right)
\end{eqnarray}

As we argued, when $T=0$ there is a closed bubble for all $K$.
Accordingly, $R$ will be analogous to the radius of this bubble.  We
define here the new dimensionless quantities $\hat{K}=\frac{K}{R^2}$ and
$\hat{p}=pR=2(1+\hat{K})$.  Our previous choice for dimensionless
quantities had the advantage of removing $p$, but had the disadvantage
that the $K\rightarrow\infty$ limit is obscured; this turns out to be
an important limit for dynamics.   Our choice here represents a change
in viewpoint.  Instead of viewing $p$ as a given parameter that we can
remove via scaling, $p$ is now a quantity which we calculate given 
$R$ and $K$.  These correspond to constant pressure versus constant 
size approaches, respectively.
\begin{figure}
\resizebox{8 cm}{!}
{
  \includegraphics{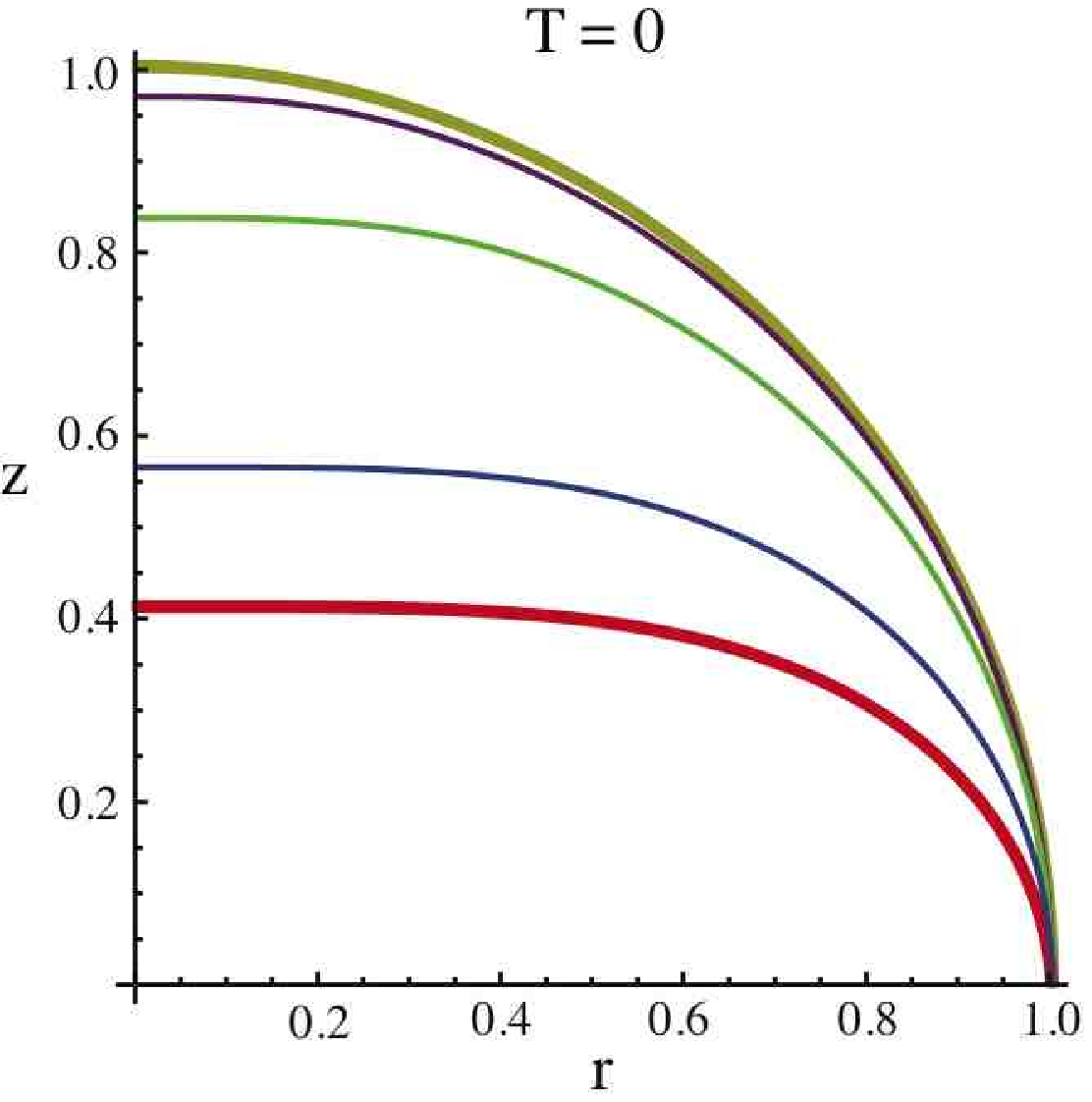}
}
\caption{Profile of free bubble ($T=0$) for $\hat{K}=0,0.01,0.1,1,
\infty$ from top to bottom.  The limiting scale-invariant
$\hat{K}=0$ (gold) and $\hat{K}=\infty$ (red) are emphasized in
bold.}
\label{figure18}       
\end{figure}
In these units, the first integral (\ref{eq:thisone}) determines the
bubble shape via:
\begin{eqnarray}
\frac{2z}{R}&=&\int_{\frac{r^2}{R^2}}^{1}\frac{\hat{p}udu}
{\sqrt{4(u+\hat{K})^2-\hat{p}^2u^3}}\nonumber\\
&=&\int_{\frac{r^2}{R^2}}^{1}\frac{(1+\hat{K})udu}
{\sqrt{(u+\hat{K})^2-(1+\hat{K})^2u^3}}
\end{eqnarray}
We can factor the cubic in the radical to find that the only real root
is at $u=1$.  This is consistent with our more general arguments (Fig.
\ref{figure11}); when
there are three critical points in $T$, $T=0$ lies in the large bubble
region, while when there is one critical point $T_1<0$, $0=T>T_1$ is
in the closed bubble region.  Written in terms of the elliptic 
functions $F$ and $E$, we have:
\begin{eqnarray}
\frac{z(r)}{R}&=&\frac{u_-}{\sqrt{1-u_-}}F\left(\sqrt{\frac{1-
\frac{r^2}{R^2}}{1-u_+}},\sqrt{\frac{1-u_+}{1-u_-}}\right)\nonumber\\
&&\qquad+\sqrt{1-u_-}E\left(\sqrt{\frac{1-\frac{r^2}{R^2}}{1-u_+}},
\sqrt{\frac{1-u_+}{1-u_-}}\right)\\
u_\pm&=&\frac{-\hat{K}}{2(1+\hat{K})^2}(2+\hat{K}\pm
i\sqrt{4\hat{K}+3\hat{K}^2})\nonumber
\end{eqnarray}
In Fig. \ref{figure18} we show the generating curve for several values
of $\hat{K}$.  Note that the shape of the curve becomes more oblate
as $\hat{K}$ increases.  Intuitively, this makes sense, as the
$K/r^2$ term in the energy penalizes surfaces with more material close
to the axis.

\subsection{Perturbative expansions}
\label{subsec:perturbation}

Though an expression in terms of elliptic integrals is compact and useful for plotting, we have garnered further insight into the shapes via perturbation for small $\hat K$.
Starting with
\begin{eqnarray}
\frac{2z}{R}&=&\int_{\frac{r^2}{R^2}}^{1}\frac{\hat{p}udu}
{\sqrt{4(u+\hat{K})^2-\hat{p}^2u^3}}
\end{eqnarray}
we expand to find this to second order in $\hat{K}$:
\begin{eqnarray}
\frac{2z}{R}&=&2\sqrt{1-\frac{r^2}{R^2}}
-\hat{K}\left(2\ln\left[\frac{R}{r}+\sqrt{
\frac{R^2}{r^2}-1}\right]\right)\nonumber\\
&&\qquad+\hat{K}^2\left(4\ln\left[\frac{R}{r}+\sqrt{
\frac{R^2}{r^2}-1}\right]+\sqrt{\frac{R^2}{r^2}-1}\right)\nonumber\\
&&\qquad+\mathcal{O}(\hat{K}^3)
\end{eqnarray}
Away from $r=0$, the first correction is negative, confirming that
the surfaces flatten
out with increasing $\hat{K}$.  Note that the terms of order
$\mathcal{O}(\hat{K})$ and above diverge at $r=0$, indicating that
the expansion is not uniformly convergent for all $r$.  Examining
(\ref{eqn:forceode2}), we see that the $\hat{K}\rightarrow 0$ limit
is singular at $r=0$ since the term with the highest power of $1/r$
drops out.

The $\hat{K}\rightarrow\infty$ limit does not suffer from
this convergence issue.  Expanding in powers of
$\kappa\equiv\hat{K}^{-1}$, we find
\begin{eqnarray}
\frac{2z}{R}&=&\int_{\frac{r^2}{R^2}}^{1}\frac{udu}
{\sqrt{1-u^3}}+\kappa\int_{\frac{r^2}{R^2}}^{1}\frac{u(1-u)du}
{(1-u^3)^{3/2}}\nonumber\\
&&\qquad+\kappa^2\int_{\frac{r^2}{R^2}}^1\frac{u^2(1-u)^2(-2-2u+u^2)du}
{2(1-u^3)^{5/2}}\nonumber\\
&&\qquad+\mathcal{O}(\kappa^3)
\end{eqnarray}
This expansion is regular for all $0\leq r\leq R$ for the
first few terms, and one can argue that this perturbation series is regular 
in $\kappa$ and $r$ since the singular point at $\hat{K}=0,r=0$ has now
moved to $\kappa=\infty,r=0$.  Fortunately, these integrals are
related to known functions (useful for plotting!):
\begin{eqnarray}
\frac{2z}{R}&=&\frac{2\sqrt{\pi}\Gamma\left(\frac{2}{3}\right)}
{\Gamma\left(\frac{1}{6}\right)}-\frac{1}{3}\text{B}\left(\left(\frac{r}
{R}\right)^6; \frac{2}{3},\frac{1}{2}\right)\nonumber\\
&&\qquad+\kappa\biggl[\frac{2}{3\sqrt{1-\left(\frac{r}{R}\right)^6}}
-\frac{4i\sqrt{\pi}\Gamma\left(\frac{11}{6}\right)}
{\Gamma\left(\frac{1}{3}\right)}\nonumber\\
&&\qquad-\frac{2i}{5}\left(\frac{R}{r}\right)^5
{}_2F_1\left(\frac{5}{6},\frac{3}{2};\frac{11}{6};\left(\frac{R}{r}
\right)^6\right)\biggr]\nonumber\\
&&\qquad+\mathcal{O}(\kappa^2)
\end{eqnarray}
where B$(z;a,b)$ denotes the incomplete Beta function and $_2F_1(a,b;c;z)$
denotes the standard hypergeometric series.  The first term gives us the
$\hat{K}\rightarrow\infty$ bubble shape, in other words the shape of
a bubble with only nematic elastic energy and no surface tension
energy.  

We have used these expansions to study the shape near the north and
south poles at $r=0$ and the equator at $r=R$.  When $\hat{K}=0$ the
generating curve is a circle, quadratic near the poles and with a 
square-root cusp at the equator.  However, for $\hat{K}\neq0$, the 
behavior is quartic
at $r=0$: $z(r)=\frac{K+R^2}{4KR^3}r^4-\mathcal{O}(r^6)$.  The power
of $K$ in the denominator in the coefficient of the leading term
explains why the expansion in powers of $K$ is singular at $r=0$.
Near $r=R$, $z(r)$ is still a square root:
$z(R-\ell)=\frac{(K+R^2)\sqrt{2R}}{\sqrt{3K^2+4KR^2+R^4}}\sqrt{\ell}
-\mathcal{O}(\ell^{3/2})$.  Thus, both expansions are valid there.  

\section{Conclusion}
In the future, we will consider the dynamics of the two bubbles shown
in Fig. \ref{figure01}.  The two limits we studied here
perturbatively control the two long-time, scale-invariant limits of
the solutions.  As discussed earlier, at finite $K$, the problem has
two length scales, $R$ and $\sqrt{K}$ - as the value of
$\hat{K}=\left(\frac{\sqrt{K}}{R}\right)^2$ changes, one or the
other limit becomes more appropriate, thus as the length scale set by
geometry $R$ ({\sl e.g.} by assuming an initial volume) becomes much
smaller than $\sqrt{K}$ through diffusion, we go to the
$\hat{K}=\infty$ or ``pure nematic'' limit.

We were able to analyze this special case where the director was 
azimuthal and found the complete range of allowed shapes and topologies.  
Whether we can extend our method of analysis for longitudinal alignment 
or general alignment is an open question.

\section*{Acknowledgments}
It is a pleasure to acknowledge useful and stimulating discussions 
with R.B. Kusner and V. Vitelli.  This work was 
supported by NSF Grant DMR05-47230 and by gifts from L.J. Bernstein and 
H.H. Coburn.

\end{document}